\documentclass[manuscript]{aastex}

\slugcomment{submitted to {\it The Astrophysical Journal}}

\shorttitle{Feedback in AGN. II.}
\shortauthors{Crenshaw et al.}

\begin{document}


\title{Feedback from Mass Outflows in Nearby Active Galactic Nuclei. \\
    II. Outflows in the Narrow-Line Region of NGC~4151}


\author{D. Michael Crenshaw, Travis C. Fischer}
\affil{Department of Physics and Astronomy, Georgia State University,
25 Park Place, Suite 605, Atlanta, GA 30303, USA}
\email{crenshaw@astro.gsu.edu, fischer@astro.gsu.edu}

\author{Steven B. Kraemer}
\affil{Institute for Astrophysics and Computational Sciences, Department
of Physics, The Catholic University of America, Washington, DC 20064, USA}
\email{steven.b.kraemer@nasa.gov}

\and

\author{Henrique R. Schmitt}
\affil{Naval Research Laboratory, Washington, DC 20375, USA}
\email{schmitt.henrique@gmail.com}



\begin{abstract}
We present a detailed study of AGN feedback in the narrow-line region (NLR) of the Seyfert 1 galaxy NGC~4151. We illustrate the data and techniques needed to determine the mass outflow rate ($\dot{M}_{out}$) and kinetic luminosity ($L_{KE}$) of the outflowing ionized gas as a function of position in the NLR. We find that $\dot{M}_{out}$ peaks at a value of 3.0 M$_{\sun}$ yr$^{-1}$ at a distance of 70 pc from the central supermassive black hole (SMBH), which is about 10 times the outflow rate coming from inside 13 pc, and 230 times the mass accretion rate inferred from the bolometric luminosity of NGC~4151. Thus, most of the outflow must arise from ``in situ'' acceleration of ambient gas throughout the NLR. $L_{KE}$ peaks at 90 pc and drops rapidly thereafter, indicating that most of the kinetic energy is deposited within about 100 pc from the SMBH. Both values exceed the $\dot{M}_{out}$ and $L_{KE}$ determined for the UV/X-ray absorber outflows in NGC~4151, indicating the importance of NLR outflows in providing feedback on scales where circumnuclear star formation and bulge growth occur.

\end{abstract}





\section{Introduction}

Active Galactic Nuclei (AGN) provide feedback to their environments through a variety of means including direct radiation \citep{cio10}, relativistic  plasma jets \citep{fab12}, and mass outflows of ionized gas \citep{cre03}.
This feedback impacts material on a variety of size scales, from gas in the immediate vicinity (light hours) of the central supermassive black hole (SMBH) \citep{tom13} to the intergalactic and/or intracluster medium on scales of megaparsecs \citep{mal13}.
The details of energy and momentum transfer to the AGN surroundings are largely unknown, and yet they are vital for understanding how feedback might affect the formation of large-scale structure in the early Universe \citep{sca04,dim05}, chemical enrichment of the intergalactic medium \citep{kha08}, and self-regulation of SMBH and galactic bulge growth \citep{hop05}.

In this series of papers, we tackle an important piece of this puzzle, which is the significance of feedback in radio-quiet AGN, which lack strong relativistic jets and comprise 90 -- 95\% of the AGN population \citep{raf09}. Our initial concern is to quantify the feedback arising from mass outflows of ionized gas (often called ``AGN winds'') by determining their mass outflow rates ($\dot{M}_{out}$) and kinetic luminosities ($L_{KE} = \onehalf\dot{M}_{out}v^2$), and comparing these quantities to their accretion rates ($\dot{M}_{acc}$) and radiative bolometric luminosities ($L_{bol}$). We rely primarily on observations of nearby, bright AGN made at high spectral and/or spatial resolutions in order to study the physical mechanisms of AGN feedback. In particular, we concentrate on Seyfert galaxies with redshifts $z < 0.035$ and bolometric luminosities L$_{bol} =$ 10$^{43}$ to 10$^{45}$ erg s$^{-1}$ to gain insight into the importance of feedback from moderate-luminosity AGN.

In \citet[][hereafter Paper I]{cre12}, we concentrated on AGN winds in the form of outflowing UV and X-ray absorbers, detected as multiple kinematic components of blueshifted absorption lines (e.g., from C~IV and N~V in the UV, O~VII and O~VIII in X-rays). These absorbers are found at distances between 0.01 and 100 pc of the central SMBH, and have outflow velocities up to about 2000 km s$^{-1}$. In Paper I, we determined that at least 5 of the 6 Seyfert 1 galaxies with luminosities in the above range have mass outflow
rates that are 10 -- 1000 times the mass accretion rates needed to generate
their observed luminosities, indicating that most of these outflows originate
from outside their inner accretion disks. These 5 Seyferts have $L_{KE} \geq 0.1\% L_{bol}$ and at least 3 of the 5 have $L_{KE}$ in the range 0.5 -- 5\% $L_{bol}$, which 
is the range typically required by feedback models for efficient self-regulation of
black-hole and galactic bulge growth \citep[and references therein]{hop10}. We concluded that outflowing UV and X-ray absorbers in moderate-luminosity AGN have the potential to deliver significant feedback to their environments.

Another form of AGN winds has been revealed through observations of the emission-line regions in Seyfert 1 and 2 galaxies at high angular resolution ($\sim$0\farcs1) with the {\it Hubble Space Telescope} ({\it HST}) \citep{fis13,fis14} and large ground-based telescopes with adaptive optics such as {\it Gemini North} \citep[e.g.,][]{sto10}. These observations provide strong evidence for mass outflows of ionized gas in the narrow-line regions (NLRs) of nearby AGN at distances of hundreds of parsecs from the central SMBH, significantly larger than those of typical UV and X-ray absorbers (Paper I). Similar to the absorber outflows, the NLR outflows reach velocities up to $\sim$2000 km s$^{-1}$ \citep{cre05,fis13,fis14}. These outflows are of great interest because they 1) provide an opportunity to spatially resolve the mechanisms of feedback, 2) could have mass outflow rates and kinetic luminosities comparable to or greater than those of the UV and X-ray absorbers, and 3) occur on scales where much of the circumnuclear star formation takes place, indicating their potential importance for the regulation of black hole and bulge growth.

We have initiated a program to determine mass outflow rates and kinetic luminosities of NLR outflows in AGN as a function of distance from their central SMBHs. These parameters, combined with those from UV and X-ray absorbers in Paper I, should help us to evaluate the overall importance of AGN winds as sources of feedback in moderate-luminosity AGN. These studies should also help us gain a better understanding of the mechanisms of feedback in more luminous AGN at higher redshifts.

In this paper, we show that an accurate determination of $\dot{M}_{out}$ and $L_{KE}$ as a function of position in the NLR requires detailed analyses of spatially-resolved spectra and emission-line images, photoionization models of the emission-line ratios, and kinematic models of the outflows. We therefore concentrate on existing observations of a single AGN, the Seyfert 1 galaxy NGC~4151 ($z = 0.00332$), in this paper to test the feasibility of this type of study, illustrate the data and techniques that are required, and assess the need for a similar study with a larger sample.

\section{Techniques}
We rely on {\it HST} archive data obtained by ourselves and others, as well as some of our previous studies of NGC~4151, to carry out this study. Specifically, to determine $\dot{M}_{out}$ and $L_{KE}$ as a function of position in the NLR of NGC~4151, we require a kinematic model of the NLR to determine the dependence of outflow velocity of the emission-line clouds on radial distance from the SMBH, photoionization models of the emission-line ratios to determine the dependence of mass of the ionized gas on distance from the SMBH, and images of the ionized gas in the NLR to determine the total mass-loss rate as a function of position.

\subsection{Kinematics of the NLR}
In \citet{cre00}, we presented a kinematic model for the NLR of NGC~4151 based on long-slit spectra of the bright [O~III] $\lambda$5007 emission line with the {\it HST} Space Telescope Imaging Spectrograph (STIS). We obtained the spectra at two position angles with the G430L grating at a spectral resolving power of $R \approx \lambda/\Delta\lambda \approx 900$ and angular resolution of $\sim0\farcs1$. We concluded that radial outflow through a biconical structure that is hollow along its axis provides a good fit to the observed velocities of the emission-line clouds.

In \citet{das05}, we presented a follow-up study using STIS long-slit spectra at five parallel positions with the G430M grating ($R \approx 9000$), as shown in Figure 1a, to cover the majority of the NLR at higher spectral resolution. We show the biconical outflow model that matches the observations in Figure 1b. This model provides an excellent fit to the measured radial velocities of the emission-line clouds in all slit positions \citep{das05}, including the one shown in Figure 1c. The kinematic model allows us to determine the inclination of the outflow axis with respect to our line of sight (in this case 45\arcdeg) as well the geometry of the NLR with respect to the plane of the host galaxy, as shown in Figure 1d \citep[see also][]{cre10}.

For the purposes of this paper, the most important result from the above model is the determination of the true velocity law $v(r)$, corrected for projection effects, as a function of radial distance. For NGC~4151, we found that $v(r)$ increases linearly from zero within 6 pc of the SMBH to 800 km s$^{-1}$ at a turnover distance of 96 pc, followed by a linear decrease to zero (in the rest frame of the host galaxy) at a distance of 288 pc from the SMBH \citep{das05}. This velocity law is reflected in the kinematic model in Figure 1b (along with projection effects) and the model envelope for one slit position plotted in Figure 1c. We found that this velocity law, of the form $v(r) = kr^n$ at $r \leq r_t$ (turnover radius) and $v(r) = v_{max} - kr^n$ at $r > r_t$ where $n = 1$, works better than other laws with $n \neq 1$ \citep{das05}. We also found the same dependence of velocity on distance in 16 other AGN with clear signatures of radial outflow \citep{fis13}.

\subsection{Physical Conditions in the NLR}
In \citet{kra00}, we used STIS G140L, G230L, G430L, and G750L long-slit spectra of NGC~4151 to determine the reddening and physical conditions in its NLR gas along a position angle of 221\arcdeg. The spectra cover the 1500 -- 10,270 \AA\ range at a spectral resolution of $R = 500 - 1000$ in 0\farcs2$\times$0\farcs1 or 0\farcs4$\times$0\farcs1 bins. These spectra allowed us to determine the reddening at each position via the observed He~II $\lambda$1640/He~II $\lambda$4686 ratio \citep{sea78} and provided a multitude of emission lines spanning a wide range in ionization for comparison with photoionization models. We generated multi-component models at each position to successfully match the dereddened line ratios and thereby determine the physical conditions, including the ionization parameter ($U$), hydrogen number density ($n_H$), and column density ($N_H$) for each component.

An important result from the above is that we were able to determine an average density law n$_H(r)$ within the emission-line clouds as a function of distance ($r$) from the SMBH. As shown in Figure 4 of \citet{kra00}, we found a well-behaved power-law dependence in the NLR of NGC~4151: $n_H \propto r^{-n}$, with n $= 1.6\pm0.1$ and 1.7$\pm0.1$ in the SW and NE directions along the slit, respectively. We therefore adopt a powerlaw with $n = 1.65\pm0.15$ for this study.

For this paper, we went back to our models to determine the mass of ionized gas ($M_{ion}$) in each long-slit bin described above using the following equation:
\begin{equation}
 M_{ion} = N_H~\mu~m_p~L(H\beta)/F(H\beta),
\end{equation}
where the ratio of the H$\beta$ luminosity [$L(H\beta$)] to the sum of the H$\beta$ fluxes from the emitting surfaces of the model components [$F(H\beta$)] gives the total area of the emitting surfaces in that bin, $\mu$ is the mean atomic mass per proton ($=$ 1.4 for solar abundances), and $m_p$ is the proton mass.

\subsection{Mass of the Ionized Gas in the NLR}

Further analysis of the STIS long-slit spectra plus {\it HST} [O~III] images provide us with a means to determine the mass of ionized gas outside of the STIS slit. $M_{ion}$ can be expressed as a function of H$\beta$ luminosity [$L(H\beta$)] and electron density n$_e$ as: $M_{ion} \propto L(H\beta)/n_e$ \citep{pet97,ost06}. For NGC~4151, the [O~III]/H$\beta$ flux ratio is the same across the NLR to within a factor of 2 \citep{kra00} and $n_H \approx 0.85 n_e$. Given that all parts of the NLR in NGC~4151 are approximately the same distance from us, we can assume $f[O~III] \propto L[O~III]$ and express the mass of ionized gas as:
\begin{equation}
M_{ion} = s~f[O~III]/n_H,
\end{equation}
where $f$[O~III] is the measured flux of the [O~III] $\lambda$5007 emission from any portion of the NLR and $s$ is a scale factor that should be approximately the same from one location to the next.

To determine the scale factor $s$ and estimate its uncertainty due to the above approximations, we used the model-determined masses, [O~III] fluxes, and densities (n$_H$) from the bins in the long-slit data in equation (2). In Figure 2, we show $s$ as a function of position along the slit, with uncertainties calculated from those in the above parameters. The dispersion in $s$ can be traced to the assumption that all of the NLR gas has the same [O~III]/H$\beta$ ratio, whereas it can be seen in Kraemer et al. (2000) that this ratio varies by a factor of 2, so that the two points in the far SW with the most discrepant values are also the two that have the lowest [O~III]/H$\beta$ ratios. For our calculations, the average value is $s = 1.70 \pm0.63 \times 10^{20}$ M$_{\sun}~$cm$^{-1}$~erg$^{-1}$s.

To account for [O~III] emission outside of the slit, we use equation (2) and our analysis of the {\it HST} WFPC2 [O~III] images of NGC~4151 in \citet{kra08} to get the total mass of ionized gas as a function of position in the NLR. In Kraemer et al., we divided the [O~III] image into sections of elliptical annuli, as shown in Figure 3, and measured the [O~III] flux in each section. The annuli have the same position angle (60\degr) and axial ratio (0.707) as the NLR bicone shown in Figure 1  \citep[inclined 45\degr\ to our line of sight][]{das05}, so that sections along each annulus should be at similar distances from the central SMBH. The width of the annuli in the radial direction along the major axis is 3 pixels, which corresponds to 0\farcs133 for the WFPC2 PC camera or 8.5 pc at the adopted distance of NGC~4151 (13.3 Mpc, corresponding to 64 pc/$''$).

Using equation (2), the scale factor $s$, measured [O~III] flux, and adopted density law, we calculate the mass of ionized gas in each section of each elliptical annulus. We assume that $n_H(r)$ is the same in all directions, and sum the sections along each elliptical annulus to obtain $M_{ion}$ as a function of distance from the SMBH in $\Delta r = 8.5$ pc intervals. Summing the masses in all of the annuli, we find that the total mass of ionized gas in the NLR within the outer ellipse in Figure 3 is 3.0 ($\pm{0.2}$) $\times 10^5$ M$_{\sun}$.

\section{Results: $\dot{M}_{out}$ and $L_{KE}$ as Functions of Position}

We obtain the mass outflow rate $\dot{M}_{out}$ of ionized gas as a function of distance from the SMBH by dividing $M_{ion}$ for each annulus by its crossing time. Thus, $\dot{M}_{out} = M_{ion}v(r)/\Delta r$ and the associated kinetic luminosity at each position is simply $L_{KE} = \onehalf\dot{M}_{out}v^2$. We assume that the velocity law $v(r)$ holds for all locations in the NLR including those outside of the slit positions in Figure 1.

In Figure 4, we show the mass outflow rate in the NLR of NGC~4151 as a function of position. $\dot{M}_{out}$ shows a strong continuous increase up to a value of 3.0 M$_{\sun}$ yr$^{-1}$ at a distance of 70 pc, followed by decrease to 0.6 M$_{\sun}$ yr$^{-1}$ at 135 pc (the limit of the measured [O~III] fluxes in \citet{kra08}). These values are consistent with an estimate of $\sim$1.2 M$_{\sun}$ yr$^{-1}$ for the overall mass outflow rate of ionized gas in the NLR of NGC~4151 by \citet{sto10}, obtained from observations with the {\it Gemini} Near-infrared Integral Field Spectrograph (NIFS). For comparison, we show the outflow rate that would be obtained if we assumed that the mass from inside 12.8 pc moved outward according to the same velocity law (a ``nuclear outflow''), such that no additional mass is added after that point. In this case, $\dot{M}_{out}$ peaks at a value of only $\sim$0.3 M$_{\sun}$ yr$^{-1}$ at the velocity law turnover of 96 pc, which is only 10\% of the true $\dot{M}_{out}$ at its peak.

It is clear from these calculations that the mass outflow rate in the NLR of NGC~4151 is much higher than that expected from a nuclear outflow, indicating that most of the outflowing ionized gas comes from distances greater than 12.8 pc from the central SMBH. As discussed above, this provides strong evidence for in situ acceleration of ionized gas in the NLR itself, up to a distance of at least 70 pc. The decrease in $\dot{M}_{out}$ at larger distances indicates that at least some of the outflow is slowing down or stopping, possibly through interactions with ambient material, which we had originally suggested to explain the velocity turnover \citep{cre00,das05}. 

In Figure 5, we show the associated kinetic luminosity as a function of position in the NLR. $L_{KE}$ climbs rapidly as more mass is brought into the NLR outflow and more energy responsible for the acceleration of the gas is converted into kinetic energy. $L_{KE}$ peaks at a value 4.3 $\times 10^{41}$ erg s$^{-1}$ at a distance of 90 pc, close to that of the velocity turnover, followed by a steep drop. The range of $L_{KE}$ values in Figure 5 is again consistent with the overall value of 2.4 $\times 10^{41}$ erg s$^{-1}$ found by \citet{sto10}.

\section{Discussion}

To obtain the above results, we made several assumptions that deserve further examination. We assumed that the density law $n_H(r)$ from a low-dispersion spectrum along a position angle of 221\arcdeg\ was the same in all directions. The sides of the bicone in NGC~4151 that are nearly perpendicular to the disk (see Figure 1d) are fainter than those close to the disk \citep{das05,sto10}, suggesting that the densities could be lower in the former. However, the faintness could also be due to less overall material in the directions away from the galactic plane. Additional low-dispersion spectra and photoionization models at different position angles and/or offsets would be useful for testing the density law assumption.

We also assumed that our derived velocity law $v(r)$ holds for all locations in the NLR. However, our kinematic model assumes a simple bicone geometry with a sharp apex, whereas the observed distribution of ionized gas in Figure 1a shows a more extended nuclear region. Other, more complicated geometries, such as an hourglass shape seen in the NLR of NGC~1068 \citep{rif14} are possible. Nevertheless, it is clear from Figure 9 in \citet{das05} that most of the NLR clouds outside of the nominal bicone have similar velocities to those inside the bicone at the same distances.

\citet{sto10} present a kinematic model of NGC~4151 based on {\it Gemini} NIFS observations in which they favor a constant velocity as a function of position in the NLR, plus a component of low-velocity gas in the galactic plane. For comparison, we concentrate on their measured radial velocity centroids; the channel maps can be misleading because the NLR consists of large individual clouds of ionized gas with high velocity dispersions on the order of hundreds of km s$^{-1}$ \citep{hut99,das05}, and each velocity
channel samples not only clouds that have a velocity centroid in that channel,
but also the velocity wings from a number of other clouds. Based on the velocity centroids, we prefer our model because we can clearly see the turnover from acceleration to deceleration in the STIS data \citep[][Figure 1c,]{das05}, whereas it is less easily resolved in the {\it Gemini} NIFS data. The low-velocity gas in our model is simply attributed to the sides of the bicones close to the plane of the sky. Despite the different kinematic models, our values of $\dot{M}_{out}$ and $L_{KE}$ 
agree well with those of \citet{sto10}, because the mass estimates are similar and the velocity amplitudes are close.

Using high-resolution X-ray imaging with the {\it Chandra X-ray Observatory} ({\it CXO}), \citet{wan10,wan11a,wan11b} found soft X-ray emission coincident with the optical NLR and extending out to over 1 kpc from the nucleus. They concluded that the soft X-ray gas was likely photoionized and calculated $\dot{M}_{out} \approx$ 2 M$_{\sun}$ yr$^{-1}$ and $L_{KE} \approx$ 1.7 $\times 10^{41}$ erg s$^{-1}$, indicating yet another important component of feedback in more highly ionized gas.

\section{Conclusions}

Using spatially-resolved images and long-slit spectra from {\it HST}, we have for the first time been able to determine the mass outflow rate $\dot{M}_{out}$ and kinetic luminosity $L_{KE}$ of the ionized gas as a function of position in the NLR of an AGN. $\dot{M}_{out}$ in NGC~4151 peaks at a distance of about 70 pc from the SMBH and is about 10 times that expected from a purely nuclear outflow coming from inside 13pc. One possible explanation for the smaller values close to the nucleus is that $\dot{M}_{out}$ decreased steadily over the previous $\sim$10$^5$ yr. However, a more likely explanation is that most of the outflow originates further away from the SMBH, due to acceleration of ambient gas at these locations. The latter explanation is consistent with our conclusion in Paper I that the majority of the outflow associated with the UV/X-ray absorbers in NGC~4151 (and the other AGN) comes from outside of the inner accretion disk.

The above conclusion was based on our finding in Paper I that $\dot{M}_{out}$ for the UV/X-ray absorbers is 0.3 -- 0.7 M$_{\sun}$ yr$^{-1}$, which exceeds by a factor of 25 -- 50 the accretion rate of 0.013 M$_{\sun}$ yr$^{-1}$ needed to sustain NGC 4151's average bolometric luminosity of 7.4 $\times 10^{43}$ erg s$^{-1}$. Thus, most of the outflow must originate at larger distances or else the accretion disk would be quickly depleted. Using the same argument, the peak $\dot{M}_{out}$ in the NLR of NGC~4151 is 3.0 M$_{\sun}$ yr$^{-1}$, which is 230 times the accretion rate, providing strong evidence for ``in situ'' acceleration of gas at large distances from the central SMBH \citep[a conclusion also reached by][]{sto10}. The gas is likely accelerated via radiative driving from the central AGN \citep{das07}, which is reponsible for the ionization of the gas, although entrainment by a more highly ionized wind is another possibility \citep{sto10}.

As discussed in Paper I, the outflowing gas could be accelerated off a large reservoir of gas (e.g., the putative torus) that accumulated before the AGN ``turned on'' or it could be accelerated off the ``fueling flow'' over a range of distances. A third variation on this theme is that source of material could be cold gas that has moved into the ionizing bicone due to simple galactic rotation. Direct support for in situ acceleration comes from our STIS observations of Mrk 573 \citep{fis10}, where we see evidence for outflows accelerated off circumnuclear dust spirals that cross into the NLR bicone. The clumpiness of the outflows on large scales (parsecs) suggests episodic ejections of gas in the nuclear regions and througout the NLR as indicated by the clumpiness of the source material in Mrk 573. 

From the above numbers, the peak $\dot{M}_{out}$ in the NLR is 4 -- 10 times that of the UV/X-ray absorbers in NGC~4151, indicating its relative importance in terms of feedback. In addition, the peak $L_{KE}$ of 4.3 $\times 10^{41}$ erg s$^{-1}$ exceeds by a significant amount the range we determined for the UV/X-ray absorbers of $L_{KE} =$ 0.2 -- 1.5 $\times 10^{41}$ erg s$^{-1}$. The majority of the kinetic energy appears to be deposited about 100 pc from the SMBH, on scales that would affect circumnuclear star formation. Thus, in terms of both magnitude and scale, the NLR outflows in this AGN appear to be even more important than the absorber outflows in providing feedback to the circumnuclear regions. However, it is not yet clear where the deposited kinetic energy goes in this case, as there is no strong evidence for shock emission and the total soft X-ray luminosity in the NLR is only $\sim$1 $\times 10^{40}$ erg s$^{-1}$ \citep{wan11b}.

The combined UV/X-ray absorber and NLR outflows in NGC~4151 yield $L_{KE}$/$L_{bol}$ $\approx$ 0.6\% -- 0.8\%, which exceeds the benchmark of 0.5\% that we used in Paper I to indicate significant AGN feedback based on the theoretical results of \citet{hop10}. This value is on the low side compared to those for most of the AGN in Paper I based on the UV/X-ray absorbers alone. It would therefore be interesting to determine how much feedback is contributed by NLR outflows in these and other nearby AGN.

\acknowledgments



{\it Facilities:} \facility{HST (STIS)}, \facility{HST (WFPC2)}.

\clearpage



\begin{figure}
\epsscale{1.0}
\plotone{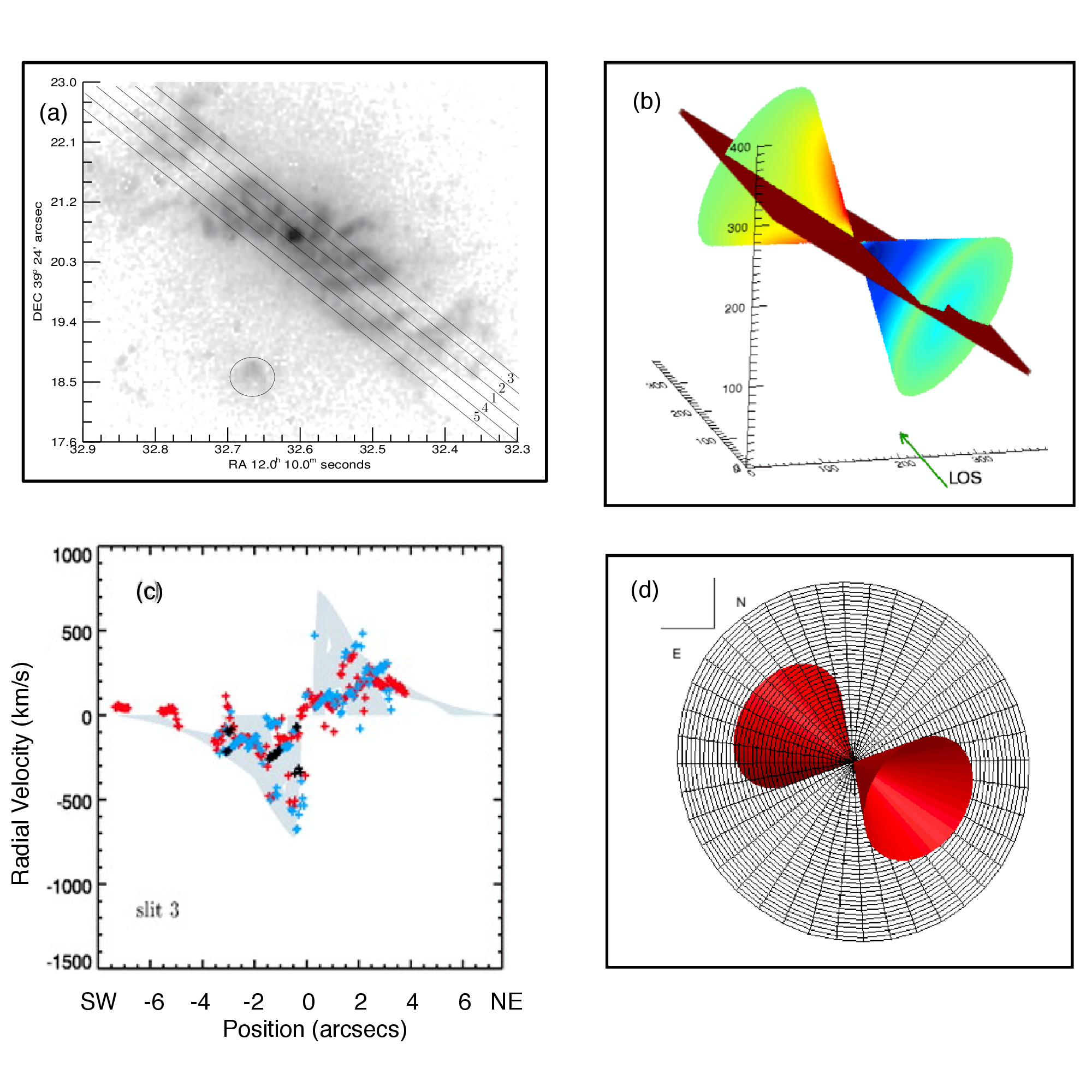}
\caption{a) [O~III] image of the NLR of NGC~4151, shown with STIS G430M slits
overlaid (the circled feature is due to a reflection in the WFPC2 camera. b) A model
representing the geometry and velocity field of the NLR of NGC~4151 is shown
with the center slit (slit 1) overlaid and the line of sight indicated (LOS) \citep{das05}.
c) Modeled radial velocities extracted from the slit are plotted in shaded gray, and compared
to the observed radial velocities of bright (red), intermediate (blue), and low (black) flux
components at each location. d) The NLR and host galaxy geometries as seen on the plane of the sky. The inclination of the bicone axis is tilted 45\arcdeg\ with respect to our LOS \citep{cre10}.\label{fig1}}
\end{figure}
\clearpage

\begin{figure}
\epsscale{1.0}
\plotone{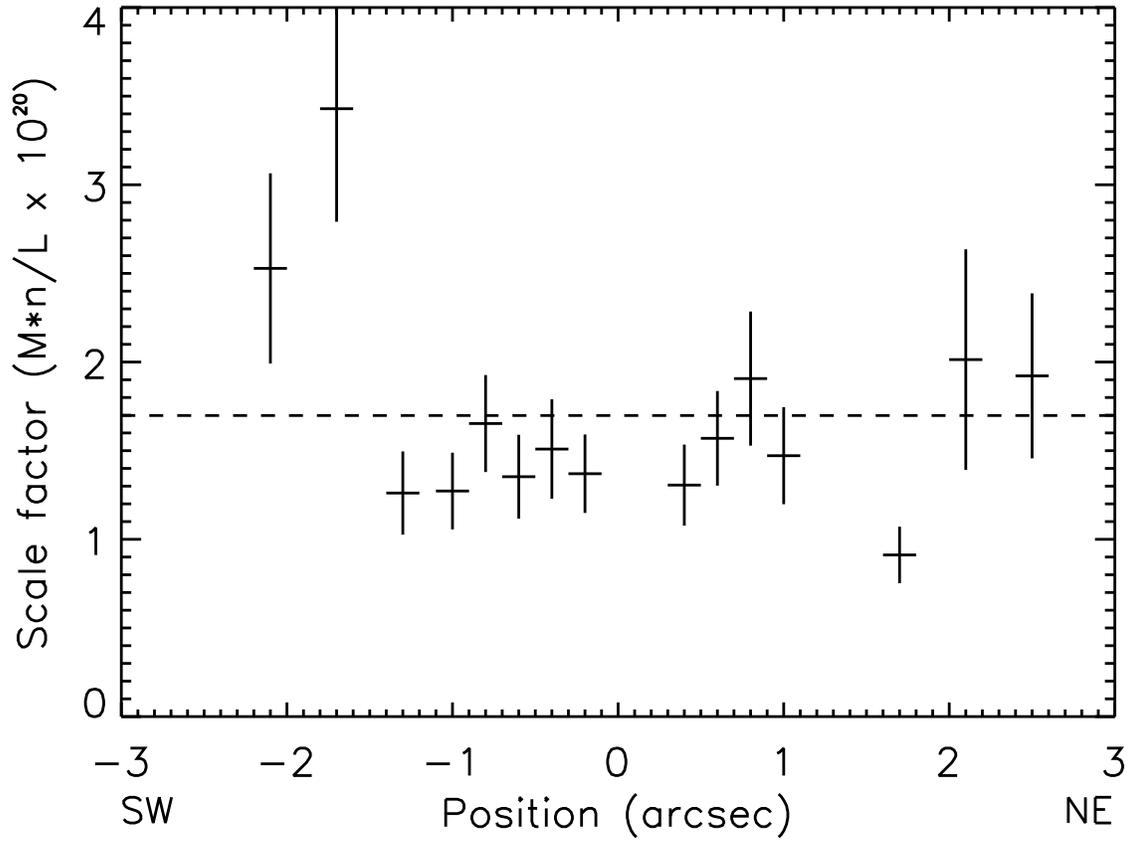}
\caption{Scale factor as a function of position in the NLR of NGC~4151 along a position angle of 221\degr (see the text). The average value is given by the dashed line. \label{fig2}}
\end{figure}
\clearpage

\begin{figure}
\epsscale{0.8}
\plotone{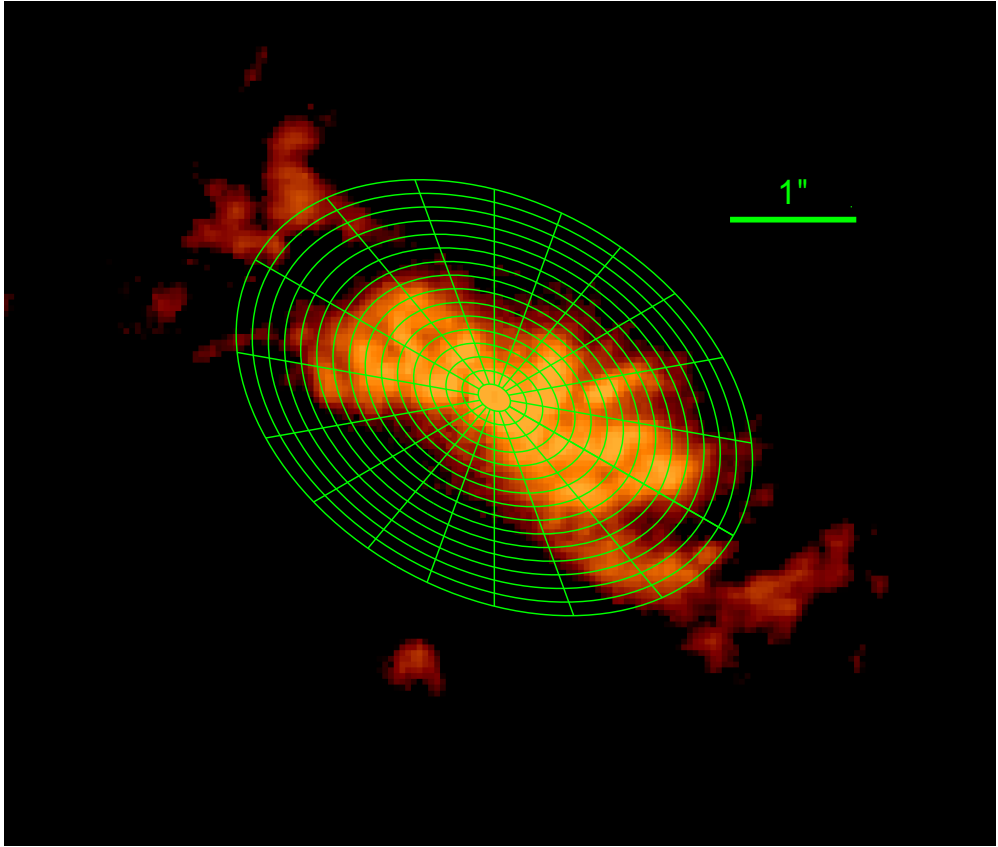}
\caption{[O~III] image of NGC 4151, showing in green the distribution
of annuli and sectors used to measure the emission-line fluxes \citep[see][]{kra08}.}
\end{figure}
\clearpage

\begin{figure}
\epsscale{1.0}
\plotone{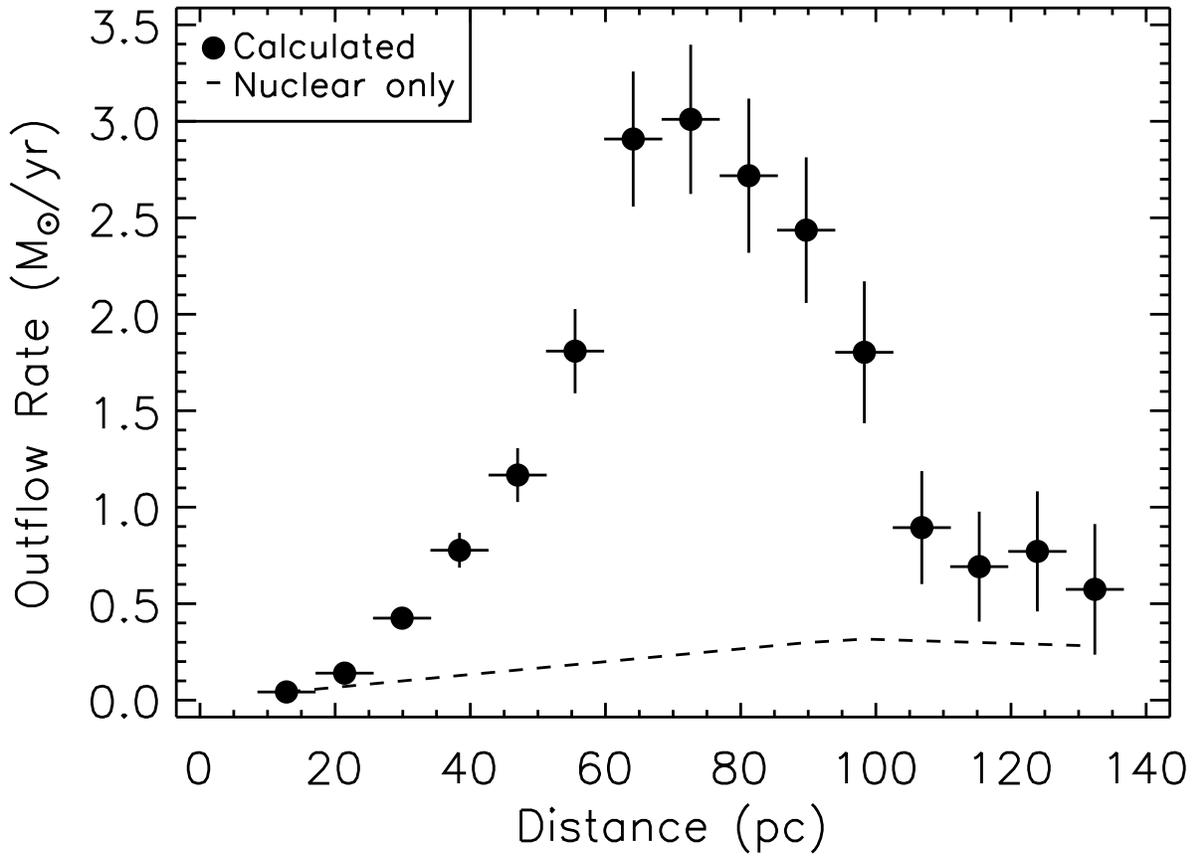}
\caption{Calculated mass outflow rate in the NLR of NGC~4151, and the rate assuming the mass originates only in the central (nuclear) bin. The error bars were propagated from the uncertainties in the parameters in equation (2).\label{fig4}}
\end{figure}
\clearpage

\begin{figure}
\epsscale{1.0}
\plotone{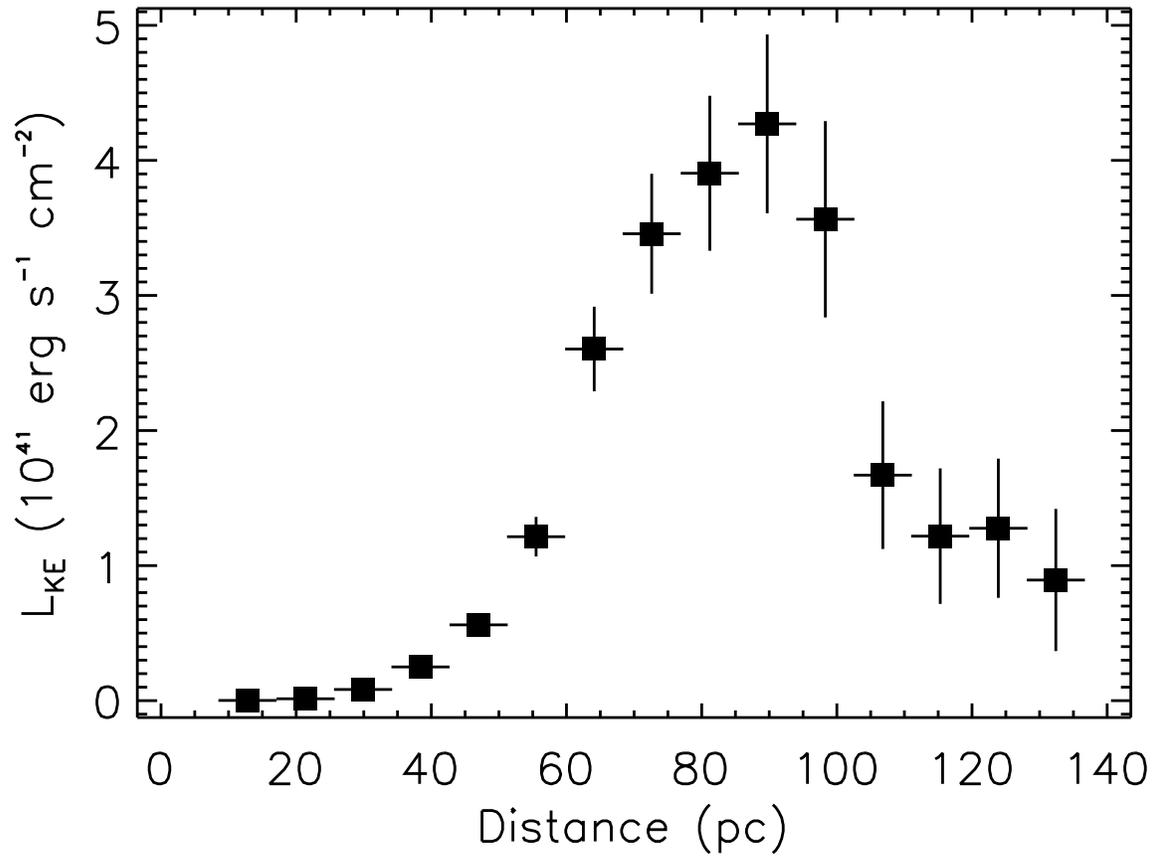}
\caption{Calculated kinetic luminosity as a function of position in the NLR of NGC~4151 Error bars were calculated as in Figure 4.\label{fig5}}
\end{figure}
\clearpage






\end{document}